\documentclass[twocolumn,secnumarabic,amssymb, nobibnotes, aps, prl]{revtex4-2}

\usepackage{graphicx}
\usepackage{dcolumn}
\usepackage{bm}
\usepackage{xcolor}

\begin{document}

\preprint{APS/123-QED}

\title{Domain Wall Reactions in Multiple-order Parameter Ferroelectrics}

\author{Songsong Zhou}
\author{Shihan Qin}%
\author{Andrew M. Rappe}%
\email{Contact author:rappe@sas.upenn.edu}
\affiliation{%
Department of Chemistry, University of Pennsylvania, Philadelphia, Pennsylvania 19104–6323, USA
}%

\date{\today}

\begin{abstract}
The motion of domain walls is crucial for ferroelectric switching. Conventionally, the switching dynamics is believed to be determined by the motion of one or a few low-energy domain wall types of dominant population. Here, we challenge this conventional idea in multiple-order-parameter ferroelectrics. Using hafnia as  example, we show that multiple-order-parameter nature not only provides various mobile domain walls and defect-like immobile domain walls, but also enables the domain wall reactions. In analogy with chemical reactions where substances react to form new substances, domain walls could also react to form other domain walls during switching. We identify several elementary domain wall reaction types including synthesis, decomposition, and exchange reactions. Domain walls are continually changed by these reactions during switching so that the switching behavior reflects the statistical average of many domain wall types with distinct mobility and stability. These reactions also lead to phenomenon like remanent nuclei and defect site nucleation that facilitate switching and lower coercive field. Finally, the concept of domain wall reaction is not limited to hafnia but can be generalized to any multiple-order-parameter ferroelectric. This work conceptualizes domain wall reaction, expands theory of ferroelectric switching, and suggests a practical way for defect engineering to control switching behavior. 
\end{abstract}

\maketitle


Domain walls (DWs), the interfaces separating different polar domains, are critical to many applications based on ferroelectrics, such as ferroelectric random access memory\cite{scott89p1400}. 
Storing data to the memory means switching the polarization direction, increasing the size of one polar region at the expense of another and causing DW motion\cite{grigoriev06p187601,ahn04p488}.
Conventional ferroelectrics such as perovskite oxides generally have only one order parameter, the polarization itself\cite{cochran59p412}. 
Thus, their DWs are just interfaces separating domains of different polar directions, such as the most common 180$^\circ$ DW of polarization reversal.
Due to the single order parameter nature, there is only one type of 180$^\circ$ DW in these ferroelectrics\cite{meyer02p104111}.
In contrast, hafnia, the most promising candidate for next-generation devices\cite{muller11p112901,boscke11p112904,olsen12p082905,cheema20p478}, involves several non-polar order parameters besides the polarization\cite{lee20p1343,zhou22peadd5953,delodovici21p064405,qi20p214108}.
Such multiple-order parameter nature leads to various types of DWs where not only polarization but also other order parameters could be reversed across the DW\cite{qi21p12538,ding20p556,choe21p8,zhou24p15251}.
In addition to the common mobile domain walls (MDWs) of polarization reversal, we emphasize a new class of immobile domain walls (IDWs).
The IDWs are interfaces separating regions of different non-polar order parameters but the same polarization direction, so that they will not move under external electric field.
These IDWs are topological defects and would commonly exist in ferroelectric thin films, as poling cannot eliminate them. 

More interestingly, the existence of the various types of IDWs and MDWs brings unexpected domain wall reactions (DW reactions) between them.
In single order parameter ferroelectrics, only a single type of MDW exists, and when these MDWs meet each other, they annihilate (signifying complete consumption of an opposite polar domain).
By contrast, in multiple-order parameter ferroelectrics, an MDW could meet with any of the IDWs/MDWs when the polarization is switched.
We propose that in analogy to chemical reactions where one set of chemical substances transforms to another, an MDW could react with other types of IDWs/MDWs to form a new set of DWs. 
Just like chemical reactions, several basic types of DW reactions could be identified, such as synthesis reactions (combining two DWs to form a new DW), decomposition reactions (splitting a DW into two different DWs), and exchange reactions (interchanges between an MDW and an IDW).

The existence of DW reactions will have a significant impact on the switching behavior.
These reactions lead to the transformation between MDW types of distinct mobility. 
Thus, the polarization switching is not dominated by the motion of a single (e.g., the most stable) DW type, but determined by an average of the motion of many MDW types, as the MDWs continually change by reactions during the switching process.
Besides, DW reactions could lead to remanent nuclei and/or IDW defect nucleation. 
These phenomena will facilitate or skip the difficult nucleation process and induce rare but highly-mobile MDWs, thus significantly speeding up the polarization switching and lowering the coercive field.
These findings highlight the importance of inducing IDWs through defect engineering as a new way to control the switching behavior, which would be critical for ferroelectrics with difficult switching and high coercive field such as hafnia.
Finally, we also note that IDWs/MDWs and DW reactions are not limited to hafnia but are possible for all ferroelectrics with more than one independent order parameter.
The concept of DW reaction opens a new field for ferroelectric studies, and this may fundamentally change our understanding of how polarization switches. 

\begin{figure}
\begin{center}
  \includegraphics[width=8.5cm]{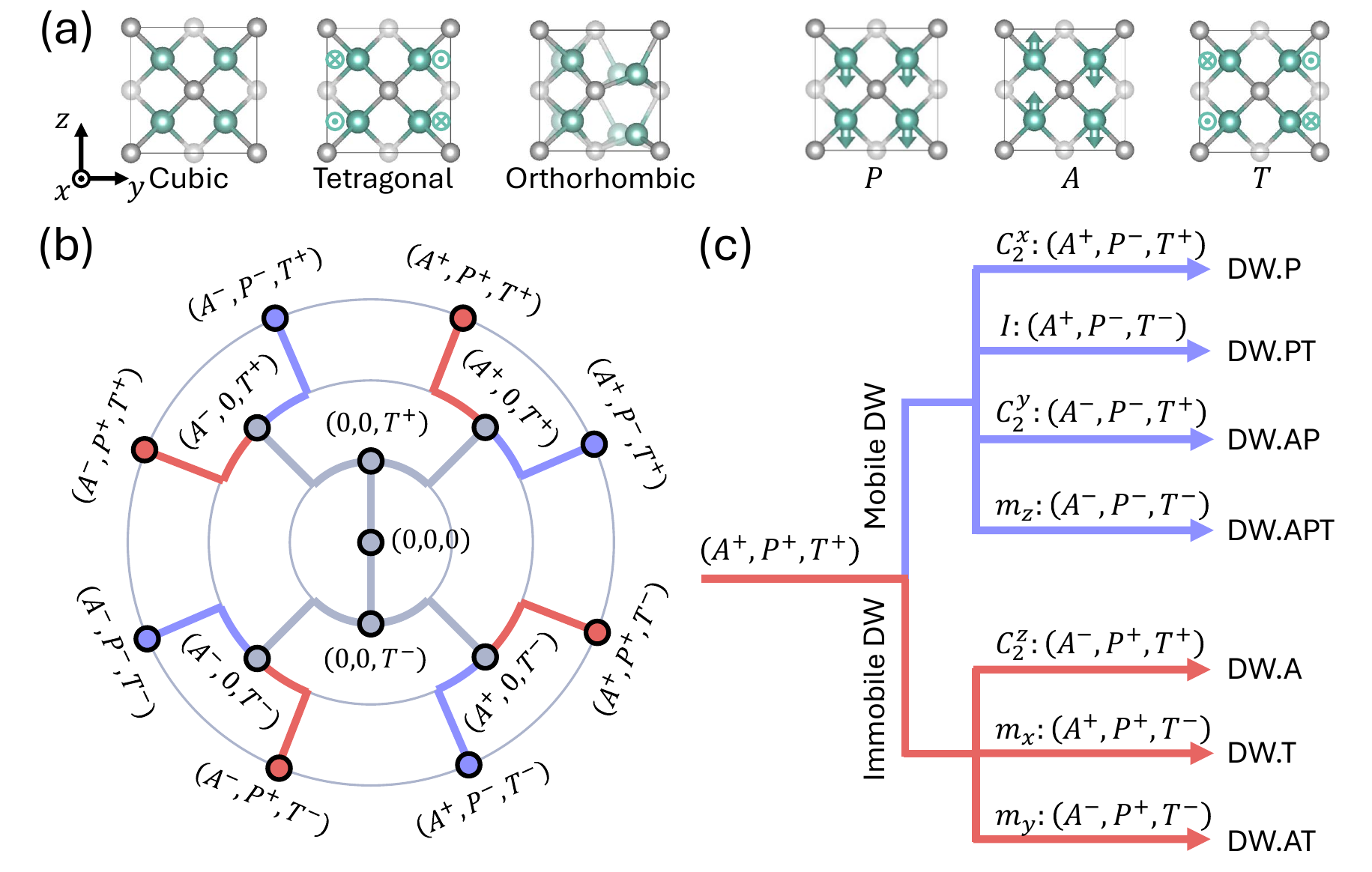} 
  \caption{(a) The structures of hafnia phases and atomic displacement of order parameters. (b) The phase transition sequence to ferroelectric orthorhombic phase. The cubic, tetragonal, and orthorhombic phases are described by order parameter vector $(0,0,0)$ (center), $(0,0,T^{\pm})$ (inner circle), and $(A^{\pm},P^{\pm},T^{\pm})$ (outer circle), respectively. Four down-polarized (blue) and four up-polarized (red) orthorhombic variants can be generated. (c) Formation of various mobile/immobile domain walls by combining symmetry-related variants of orthorhombic phase. }
  \label{figure1}
\end{center}
\end{figure}

\begin{figure*}
\begin{center}
  \includegraphics[width=17cm]{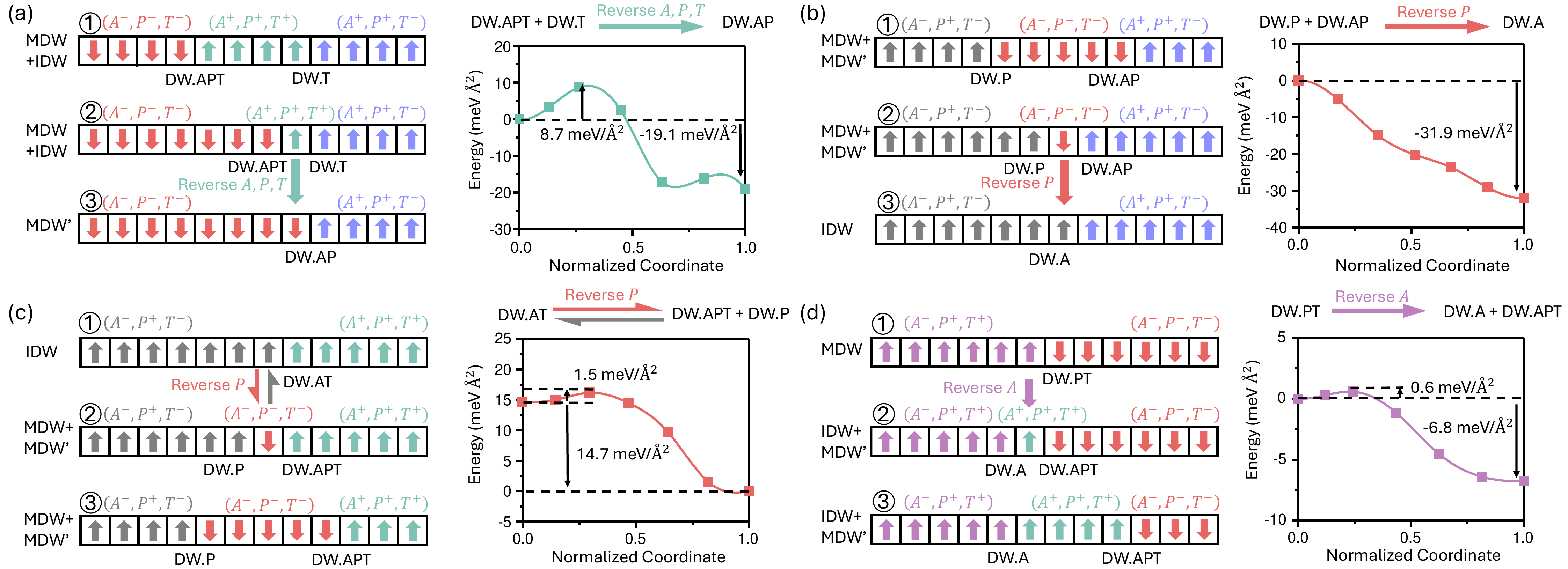}
  \caption{Schematic process and minimum energy reaction path of (a) MDW synthesis; (b) IDW synthesis; (c) IDW decomposition (reverse reaction of thermodynamically unfavorable IDW synthesis) and (d) MDW decomposition reaction.
  }
  \label{figure2}
\end{center}
\end{figure*}

Here, we use hafnia as an example to illustrate the IDWs and DW reactions.
We start with the order parameters, phases, and DW types in hafnia.
Hafnia crystallizes in the cubic fluorite structure at high temperature. Upon cooling below 2870 K, the material undergoes phase transition to the tetragonal phase by condensing the tetragonal mode (order parameter $T$)\cite{terki08p1484,wang92p5397}. 
The ferroelectric orthorhombic phase is induced by condensing three additional order parameters: polar mode ($P$), antipolar mode ($A$), and
non-polar mode ($M$)\cite{sang15p162905,lee20p1343,materlik18p164101,reyes14p140103}.
The orthorhombic phase is thus described by an order parameter vector $(A^{\pm},P^{\pm},T^{\pm})$, where the superscripts denote the sign of each order parameter (only three of four order parameters are independent, and thus $M$ is omitted, see details in SM\cite{sm}).
\nocite{kresse96p15}
\nocite{kresse96p1169}
\nocite{blochl94p17953}
\nocite{kresse99p758}
\nocite{mills95p305}
\nocite{momma11p1272}
\nocite{muller12p4318}
\nocite{li20p2000264}
\nocite{zhou15p240}
\nocite{bousquet08p732}
Accordingly, eight degenerate variants of the orthorhombic phase could be generated for one choice of tetragonal and polar axes (Fig.1(b)). 
A DW could thus be formed by combining any variant with any of the other seven variants (Fig.1(c)).
Starting from any variant, one could combine it with any of the four variants of opposite polarization $P$ to form four MDW types. 
Besides the MDWs, one could also combine one variant with any of the three variants of the same $P$ to form three defect-like IDWs. 
These DWs are named as DW.X, where X are the order parameters that reverse their signs across the DW, e.g., DW.P is the MDW where only the sign of $P$ is reversed, while the other order parameters are preserved. 
Among the four MDWs, the DW.APT is the most stable DW type, has extremely low energy but is also very hard to move\cite{lee20p1343}. 
The low mobility of DW.APT is believed to be the origin of a major outstanding issue for ferroelectric hafnia applications - the difficulty of ferroelectric switching and high coercive field\cite{wang21p010902,muller15pN30,lee19p8929,zhou15p240}.
The DW.AP and DW.P have higher DW energies than DW.APT but are easy to move, making them conducive to fast polarization switching and low coercive field (the fourth option DW.PT, as shown later, will decompose and does not play a significant role)\cite{choe21p8,qi21p12538}.

\begin{figure}[!b]
\begin{center}
  \includegraphics[width=8.5cm]{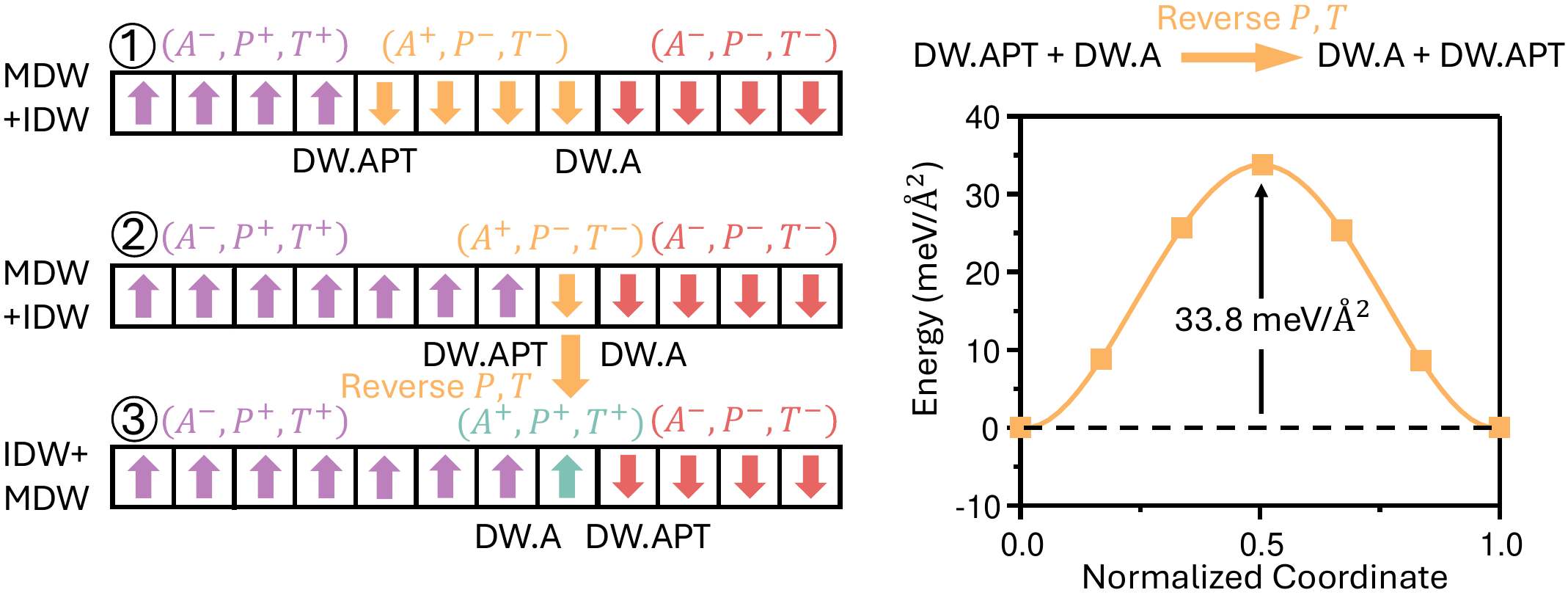}
  \caption{ Schematic process and minimum energy reaction path of the exchange reaction.}
  \label{figure3}
\end{center}
\end{figure}

Having identified the DW types for hafnia, we now investigate the basic types of DW reactions, starting from the synthesis reaction.
A synthesis reaction happens when one MDW moves toward another DW (reactants) and then merges with it to form a new DW (product). 
Such reactions could be further classified into MDW synthesis reactions (one MDW merges with an IDW into a new MDW) and IDW synthesis reactions (two different MDWs merge into an IDW).

Here, we study the combination of DW.APT and DW.T to form DW.AP as an example of the MDW synthesis reaction.
We start by modeling a thin film that contains an IDW (DW.T) and an MDW (DW.APT) that are separated by an intermediate domain $(A^+,P^+,T^+)$ (Fig.2(a),\textcircled{1}).
Under external electric field, the DW.APT moves toward the immobile DW.T by consuming the intermediate domain bit by bit. 
The DW reaction is ready to begin when sufficient intermediate domain is consumed so that the two DWs are adjacent, with only one unit cell plane of intermediate domain separating them (Fig.2(a), \textcircled{2}). 
The polarization of the last plane of $(A^+,P^+,T^+)$ will be next to reversed.
Such reversal could be completed by reversing all three order parameters so that the last plane of $(A^+,P^+,T^+)$ is converted to $(A^-,P^-,T^-)$ and the two DWs merge into a new MDW (DW.AP) (Fig.2(a), \textcircled{2}$\rightarrow$\textcircled{3}). 
The DW.AP will continue to move forward under the external field.
Through density functional theory simulation, we find that this reaction is facile from both thermodynamic and kinetic points of view (see simulation detail in SM\cite{sm}). 
The merging of DW.APT and DW.T to form DW.AP reduces the total energy of the system by 19.1 meV/{\AA}$^2$, suggesting this reaction is thermodynamically favorable (Fig.2(a)).
Kinetically, merging the two DWs needs to overcome a moderate barrier of 8.7 meV/{\AA}$^2$, significantly lower than the barrier ($>$40 meV/{\AA}$^2$) to move DW.APT\cite{lee20p1343}. 

It should be noted that the DW.APT is the most stable (and likely the most plentiful) but hard to move DW, which is therefore responsible for the slow switching and high coercive field in hafnia.
On the other hand, the DW.AP is found to be of higher energy (and thus rare) but easy to move\cite{qi21p12538, zhou24p15251}. 
Through this synthesis reaction, the stable DW.APT is converted to the higher energy DW.AP due to the elimination of the defect-like DW.T, regaining its formation energy. 
Thus, such synthesis reaction is very favorable for easier polarization switching and lower coercive field in hafnia.

Besides the MDW synthesis reaction, the IDW synthesis reaction could also happen when two different MDWs move toward each other and merge into an IDW. 
Such a reaction signifies the complete switching of a domain. 
For example, the DW.P and DW.AP could merge into DW.A by reversing the sign of $P$ in the separating domain (Fig.2(b), \textcircled{2}$\rightarrow$\textcircled{3}). 
This reaction lowers the total energy by 31.9 meV/{\AA}$^2$ and has no energy barrier (Fig.2(b)), so it is both thermodynamically and kinetically favorable. 
Such IDW synthesis reactions suggest that the IDWs will not be eliminated by MDW synthesis reactions during the switching process, as they will also be created when the domain switching completes.
Thus, MDW synthesis reactions could still happen in the subsequent switching processes. 

However, the IDW synthesis reactions are not always thermodynamically favorable as the reaction product (IDW) may not necessarily have lower energy than the reactants (two MDWs). 
For example, merging DW.APT and DW.P into DW.AT (by reversing $P$ of intermediate domain) will cost an additional energy of 14.7 meV/{\AA}$^2$ (Fig.2(c) \textcircled{2}$\rightarrow$\textcircled{1}).
Such an IDW synthesis reaction is not favorable, and the two MDWs will repel each other when they are too close. 
To merge these two MDWs, a large external electric field is needed to overcome their repulsion, as the energy gain from switching the last bit of intermediate domain would be larger than the energy cost for merging the MDWs (see SM for detail\cite{sm}).
However, if the electric field is not large enough, the two MDWs will not merge but instead leave a small unswitched intermediate domain. 
This domain is a ``remanent nucleus''.
In subsequent polarization switching events, such remanent nuclei could skip the difficult nucleation process that forms the initial reversal polar domain and MDWs.
Like a racing car with its engine on before the starting light, these remanent nuclei will grow immediately upon applying electric field, and thus significantly speed up the switching process and lower the coercive field.

Just like the relationship between forward and reverse chemical reactions, the existence of a thermodynamically unfavorable DW synthesis reaction would naturally suggest the favorability of its reverse reaction -- DW decomposition reaction.
The two types of synthesis reactions above have corresponding decomposition reactions: the IDW decomposition (decomposing an IDW into two different MDWs) and the MDW decomposition reaction (decomposing an MDW into an IDW and a new MDW). 

Returning to the above example of a thermodynamically unfavorable IDW synthesis reaction, the IDW decomposition reaction of DW.AT is hence thermodynamically favorable. 
DW.AT could be decomposed to DW.APT and DW.P by creating a new intermediate domain of reversed $P$ (Fig.2(c), \textcircled{1}$\rightarrow$\textcircled{2}). 
This process could be considered as defect-site nucleation, where the nucleation initiates from a IDW defect rather than directly nucleating within the domain.
This process needs to overcome a barrier of only 1.5 meV/{\AA}$^2$ (Fig.2(c)), which would be facile under low external electric field.
This suggests that compared to conventional bulk nucleation within a pristine domain (with a barrier $>40$ meV/{\AA}$^2$), nucleation near a DW.AT though IDW decomposition reaction would be much easier, enabling lower coercive field and faster switching.
Moreover, unlike bulk nucleation that always creates two DW.APT due to its extremely low DW energy, the defect-site nucleation here creates one DW.P besides DW.APT.
DW.P is known to be highly mobile, but its presence would be rare due to its high DW energy compared to DW.APT\cite{choe21p8}. 
Thus, the nucleation though DW.AT decomposition provides a way to solve the outstanding problem of how to induce the highly mobile but rare MDWs. 

\begin{figure}
\begin{center}
  \includegraphics[width=8.5cm]{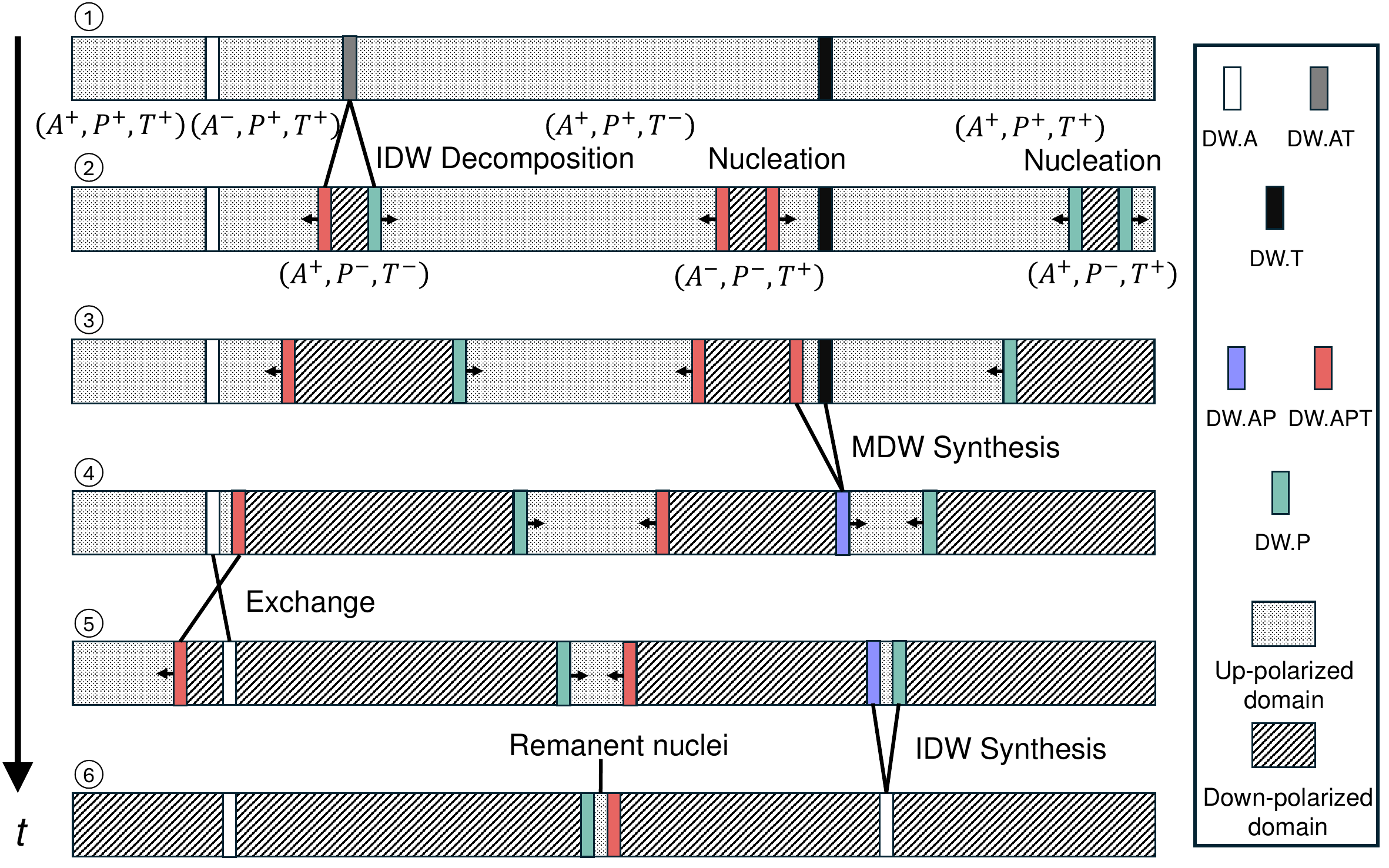}  
  \caption{Schematic examples of polarization switching process determined by various DW reaction events.  }
  \label{figure4}
\end{center}
\end{figure}

Besides the IDW decomposition, MDWs could also be decomposed, which is the reverse reaction of MDW synthesis reaction. 
For example, DW.PT may be decomposed into a DW.A and a DW.APT by generating a new intermediate domain of reversed $A$ mode (Fig.2(d), \textcircled{1}$\rightarrow$\textcircled{2}).
Although the reversal of nonpolar mode cannot be driven by external electric field, the extremely low barrier of only 0.6 meV/{\AA}$^2$ (close to thermal fluctuation energy at room temperature) ensures that the decomposition could still happen due to the thermal fluctuation (Fig.2(d)).
Such reaction would lower the total energy of the system by 6.8 meV/{\AA}$^2$.
Thus, the DW.PT is an unstable DW at room temperature, which explains why this MDW is rarely found in experimental observations. 
This dynamic DW decomposition reaction also provides a new perspective to evaluate the stability of DW structure, beyond the conventional evaluation of thermodynamic and kinetic stability of static single domain structure. 

The existence of thermodynamically favorable MDW decomposition reactions (e.g., Fig.2(d)) means that the moving MDWs may not always be able to merge with the IDW defects. 
However, it does not suggests this MDW must be pinned by the IDW defects.
When other reactions are impossible, the MDW may still pass through the IDW by exchange reaction without changing the population of DW types.
For example, the DW.APT and DW.A cannot not merge into DW.PT (Fig.2(d)). 
Instead of being pinned by DW.A, the original DW.APT can by pass the DW.A, switching the DW.A and DW.APT locations (Fig.3, \textcircled{2}$\rightarrow$\textcircled{3}).
Thus the energy of the system is not changed (i.e., always thermodynamically facile). 
Macroscopically, such exchange of position allows the MDW to pass through the IDW defect.
Whether the MDW could pass through the IDW is hence determined by its kinetic.
Here, the reaction barrier is 33.8 meV/{\AA}$^2$, smaller than the barrier for DW.ATP propagation ($>$40 meV/{\AA}$^2$).
Thus, an electric field sufficient to switch the hafnia thin film (i.e., move the DW.APT) would also be able to drive exchange reaction.
Reversely, in cases of high exchange reaction barrier, the MDW will be effectively pinned by IDW defects and jeopardize switching.

\begin{figure}[!b]
\begin{center}
  \includegraphics[width=8.5cm]{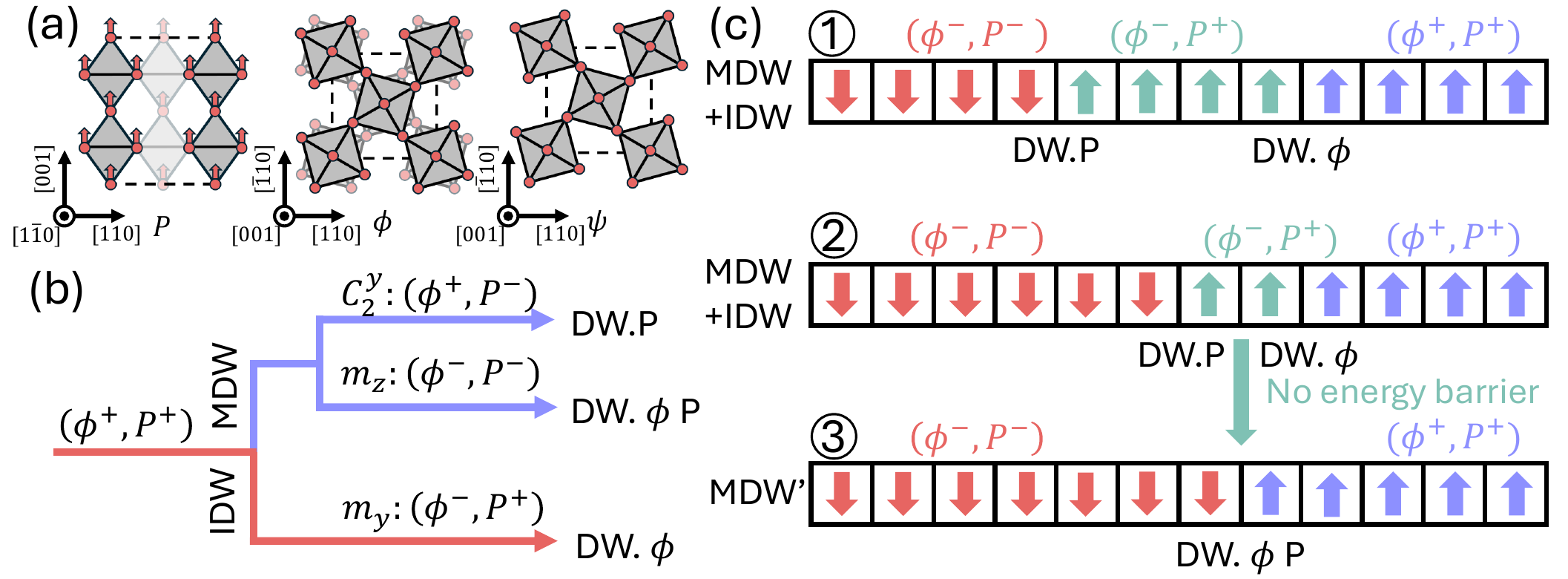} 
  \caption{(a) Schematic of structural distortion of order parameters in improper ferroelectric PbTiO$_3$/SrTiO$_3$ superlattice. Only oxygen atoms (red) and octahedral cages (grey) are shown for clarity
  (b) Two independent order parameters $(\phi,P)$ can form two MDWs and one IDW. (c) Schematic of synthesis reaction process in PbTiO$_3$/SrTiO$_3$ superlattice.}
  \label{figure5}
\end{center}
\end{figure}

In light of the various types of DW reactions, the polarization switching process in multi-order parameter ferroelectric is not dominated by one single type of DW.
Instead, such process involves many types of DWs and their reactions, such as those schematically illustrated in Fig.4.
The initial state of the ferroelectric sample contains multiple up-polarized domains as well as IDW defects separating them (Fig.4, \textcircled{1}).
Under external electric field, the switching process is initialized by nucleation, involving both direct nucleation within a domain and defect site nucleation through IDW decomposition reaction (\textcircled{1}$\rightarrow$\textcircled{2}). 
The nuclei keep growing under external electric field (\textcircled{3}).
During the growth, the MDWs could meet IDWs and change their types by MDW synthesis reactions (\textcircled{3}$\rightarrow$\textcircled{4}).
When the synthesis reaction is not facile, the MDWs may also pass through the IDWs by exchange reactions (\textcircled{4}$\rightarrow$\textcircled{5}).
Finally, the domain switching ends when the MDWs meet each other. 
This may leave an IDW defect through IDW synthesis reaction or leave a remanent nucleus for the next switching (\textcircled{5}$\rightarrow$\textcircled{6}).
Thus, in contrast to single-order parameter ferroelectric, the overall behavior of polarization switching in multi-order parameter ferroelectrics such as hafnia reflects the statistical average of the motions and reactions of many DW types. 
Moreover, the remanent nuclei and defect site nucleation through IDW decomposition reaction may also facilitate switching.

Finally, we note that DW reactions are not special cases limited only to hafnia, but a general phenomenon that could happen in any ferroelectric system with more than one independent order parameter, such as (hybrid) improper ferroelectrics\cite{benedek11p107204,levanyuk74p199,benedek22p331}.
Here, we use a PbTiO$_3$/SrTiO$_3$ superlattice as an example to prove that DW reactions could generically happen in multiple order parameter ferroelectrics.
In this perovskite superlattice, the out-of-phase ($\phi$) and in-phase tilts ($\psi$) freeze in together, breaking the inversion symmetry\cite{bousquet08p732}.
The presence of these two tilts enable a non-zero polarization ($P$) via the trilinear coupling (${\phi}{\psi}P$).
Thus, only two of the three order parameters are independent, and the domains could be described by $({\phi}^{\pm},P^{\pm})$ (omitting $\psi$ mode). 
In this system, two MDWs and one IDW could be formed (Fig.5(b)).
The DW reaction could also happen in such a system.
For example, a MDW synthesis reaction in this superlattice happens when a MDW (DW.P) approaches an IDW (DW.$\phi$) (Fig.5(c)).
Note that due to the existence of two degenerate polar layers in each unitcell (there is only one polar layer in hafnia), the DWs could move in distance of half unit cell width. Thus, one unit cell contains two blocks in Fig.5(c).
We find that when the two DWs are separated by intermediate domain of only one unit cell, the structure is destabilized so that the two DWs spontaneously merge to one single DW.$\phi$P (i.e., the synthesis reaction).
The synthesis process is completed by converting half of the intermediate domain into $({\phi}^-,P^-)$ and $({\phi}^+,P^+)$ domain, respectively (See SM for detail\cite{sm}).
Such a spontaneous process also suggests this synthesis reaction is favored thermodynamically (energy lowering) and kinetically (zero energy barrier).

In summary, our work demonstrates the existence of various MDWs/IDWs and the DW reactions that occur in hafnia and other multi-order parameter ferroelectric systems. 
These reactions not only provide a new perspective on understanding domain wall stability and kinetics, but more importantly imply a switching process involving multiple DWs and defects whose structures and populations are constantly evolving. 
This insight challenges the conventional point of view that polarization switching is determined by the motion of the dominant low-energy DW, suggesting that the switching behavior reflects the average behavior of many DW types of distinct mobility and stability.
From a practical application perspective, it also highlights the remanent nuclei and defect site nucleation through IDW decomposition reaction, which skip or facilitate the nucleation process and induce rare but desired highly mobile DWs.
These findings suggest that inducing IDW defects by defect engineering could be an effective way to increase the polarization switching speed and lower coercive field in hafnia. 

\begin{acknowledgments}
{\it{Acknowledgements}} $\--$ S.Z., and S.Q. acknowledge the support of the U. S. Department of Energy, Office of Science, Office of Basic Energy Sciences Energy Frontier Research Centers program under Award Number DE-SC0021118, for the study design and scientific insight. 
A.M.R. acknowledge of the same support for the supervision and analysis of the study. 
Computational support was provided by the High-Performance Computing Modernization Office of the Department of Defense and the National Energy Research Scientific Computing Center (NERSC), a U.S. Department of Energy, Office of Science User Facility located at Lawrence Berkeley National Laboratory, operated under Contract No.DE-AC02-05CH11231.
\end{acknowledgments}

\bibliography{apssamp}

\end{document}


\preprint{APS/123-QED}

\title{Supplemental Materials: Domain Wall Reactions in Multiple-order Parameter Ferroelectrics}

\author{Songsong Zhou}
\author{Shihan Qin}%
\author{Andrew M. Rappe}%
\affiliation{%
Department of Chemistry, University of Pennsylvania, Philadelphia, Pennsylvania 19104–6323, USA
}%

\date{\today}


\maketitle
\renewcommand{\figurename}{Fig.S}

\renewcommand*{\citenumfont}[1]{S#1}
\renewcommand*{\bibnumfmt}[1]{[S#1]}
\section{DFT Simulations} Density functional theory calculations were performed using the Vienna Ab-initio simulation package (VASP) \cite{kresse96p15,kresse96p1169} with a plane-wave basis set and the projector augmented-wave method\cite{blochl94p17953,kresse99p758}.
The exchange correlation energy functional was described by the local density approximation (LDA).
The plane-wave cutoff was set to 500 eV.
The Brillouin zone of hafnia and the perovskite superlattice unit cell were sampled by $4\times4\times4$ and $4\times4\times3$ k-point meshes, respectively.
Each domain wall structure was fully relaxed until the total force on each atom is smaller than 0.01 eV/{\AA}. 
The reaction path and energy barrier between initial and final states of domain wall reactions was calculated by the nudged elastic band (NEB) method\cite{mills95p305}.
For the reactions involving zero energy barrier, the initial state for NEB calculation was relaxed under fixed polar displacement to avoid them relaxing directly into the final state.
The figures of atomic structures and modes are plotted by the VESTA code\cite{momma11p1272}. 

\section{Phonon mode and order parameter} Based on the symmetry of cubic phase, eight modes are involved during the transition from cubic phase to orthorhombic ferroelectric phase, as illustrated in Fig.S1 \cite{zhou24p15251,delodovici21p064405,qi20p214108,reyes14p140103}.
During the first step cubic to tetragonal phase transition, the tetragonal mode ($T$) condenses in. 
Then, based on the symmetry of tetragonal phase, the remaining seven modes combine linearly with each other to form three modes (polar $P$, antipolar $A$, and nonpolar $M$ modes) that connect the tetragonal phase to the orthorhombic phase.
The antipolar mode $A$ is the linear combination of $Y_5^z$, $Z_5^{y-}$, and $Z_5^{x-}$, where the $Y_5^z$ is the major component displacement of large amplitude, while the other two modes are of very small amplitude.
Similarly, the $M$ mode is the linear combination of $Z_5^x$ and $Y_3^-$, where the first one is the major component displacement and the last one is of small amplitude.
The polar mode $P$ involves $\Gamma^z$ and $X_5^y$, where the first one involves uniform oxygen displacement along the z-axis, which is the major component for polarization.
More detail about mode linear combinations could be found in previous studies\cite{zhou24p15251}.
Note that in Fig.1 of main text, only the major components of these three modes are plotted for simplicity.
The trilinear relationship $APM$ is identified according to the symmetry of tetragonal phase; thus, once the sign of two of these order parameters is known, the sign of the other order parameter could be determined to minimize the trilinear coupling energy term $APM$\cite{delodovici21p064405}.
Here, we choose to omit the $M$ and choose $A$ and $P$ as order parameters.
In addition to the $T$ mode involved in the first step transition, a vector of three independent order parameters $(T,A,P)$ is chosen.

\section{Merging two MDWs under electric field} The total energy of the system involved in an IDW synthesis reaction can be divided into two parts: the energy difference between the two reactant MDWs and the product IDW; and the energy gain due to the polarization reversal of the intermediate domain that separates the two MDWs. 
Then the change of energy $\Delta U$ can be expressed as $\Delta U= A\Delta{\gamma}-2EPAs$,
where $\Delta{\gamma}$ is the energy difference between the product IDW and the two reactant MDWs; $A$ is the domain wall area; $E$ is the external electric field; $P$ is the polarization; $s$ is the width of the last bit of intermediate domain so that $V=As$ is the volume.
Thus, when the IDW is of higher energy than the two MDWs (i.e., $\Delta{\gamma}>0$), the critical electric field needed to drive the reaction ($\Delta U<0$) would be estimated as: $E>\Delta{\gamma}/2Ps$.
Taking $P=53$ $\mu${C}/cm$^2$ from previous work\cite{zhou22peadd5953}, $s$=5{\AA} as the width of only one unit cell, and $\Delta{\gamma}=14.7$ meV/\AA$^2$ for merging DW.APT and DW.P to DW.AT, we could estimate the critical electric field $E=4.5$ MV/cm for merging these two MDWs.
This is a quite large electric field that exceeds the field applied in most experimental studies of hafnia switching\cite{muller12p4318,li20p2000264,zhou15p240}. 

\section{Domain wall synthesis in improper ferroelectric perovskite superlattice} The unit cell of PbTiO$_3$/SrTiO$_3$ perovskite superlattice is constructed from the simple perovskite unit cell following the relationship: $\overrightarrow{a}^{'}=\overrightarrow{a}+\overrightarrow{b}$, $\overrightarrow{b}^{'}=-\overrightarrow{a}+\overrightarrow{b}$, $\overrightarrow{c}^{'}=2\overrightarrow{c}$.
Where $\overrightarrow{a}^{'}$,$\overrightarrow{b}^{'}$, $\overrightarrow{c}^{'}$ and $\overrightarrow{a}$, $\overrightarrow{b}$,$\overrightarrow{c}$ are the lattice vectors of superlattice and the simple perovskite, respectively\cite{bousquet08p732}.
In other word, the superlattice has $\sqrt{2}\times\sqrt{2}\times2$ pattern. 
Thus, each unit cell contains four oxygen octahedral cages (with Ti atom in each cage center), forming two planes of TiO$_6$ cages (top and bottom), see Fig.S2(a) (Note that the fainter cages are in further back than the solid cages).
The Pb and Sr atoms are located in the boundary between the planes of cages.
The two primary order parameters, i.e., the in-phase and out-of-phase Glazer tilts, superimpose constructively in one of the two planes and destructively in the other, which leaves only one plane of octahedral cages rotated. 
The signs of these two primary order parameters determine which cages will be rotated.
The $\sqrt{2}\times\sqrt{2}\times2$ pattern also provides two vertical polarization layers per unit cell; thus each unit cell contains two black blocks in Fig.S2 and Fig.5 of main text.
The polarization direction of each block is determined by the rotation pattern of corresponding octahedral cages, i.e., the sign of primary order parameters.

According to the trilinear relationship $\phi\psi{P}$, the $(\phi^-,P^-)$ domain (the red domain in Fig.5 of main text and Fig.S2), has $\phi<0, \psi>0$, so that the two primary order modes superimpose constructively in bottom plane, while leaving the cages on the top plane not rotated. 
The $(\phi^-,P^+)$ and $(\phi^+,P^+)$ domains (green and blue domain) has $\phi<0, \psi<0$ and $\phi>0, \psi>0$, respectively (i.e., two primary order parameters are of same sign).
Thus in these two domains, only the top plane of cages is rotated.
Accordingly, the DW.P is marked by the rotation disruption, i.e., the boundary between top plane rotation to bottom plane rotation (Fig.S2(b)). 
This is because $\phi$ does not change sign across the boundary but $P$ and $\psi$ change sign.
The IDW DW.$\phi$ is marked by the boundary where the octahedral cage accommodates contradictory rotations in the same (top) plane. 
These two DWs merge into a single DW.$\phi$P by changing the rotation pattern in the intermediate domain $(\phi^-,P^+)$. 
In the left half of $(\phi^-,P^+)$ (note each unit cell contains two polarization layers/blocks), the rotation in the top plane becomes zero, while the bottom plane becomes rotated to accommodate the $(\phi^-,P^-)$ pattern.
In the other half of the intermediate domain, the top plane rotation direction is reversed, which accommodates the pattern of the $(\phi^+,P^+)$ domain.
Consequently, the DW.P and DW.$\phi$ merge into the DW.$\phi$P located right in the center of the original intermediate domain.

\begin{figure}
\begin{center}
  \includegraphics[scale=0.5]{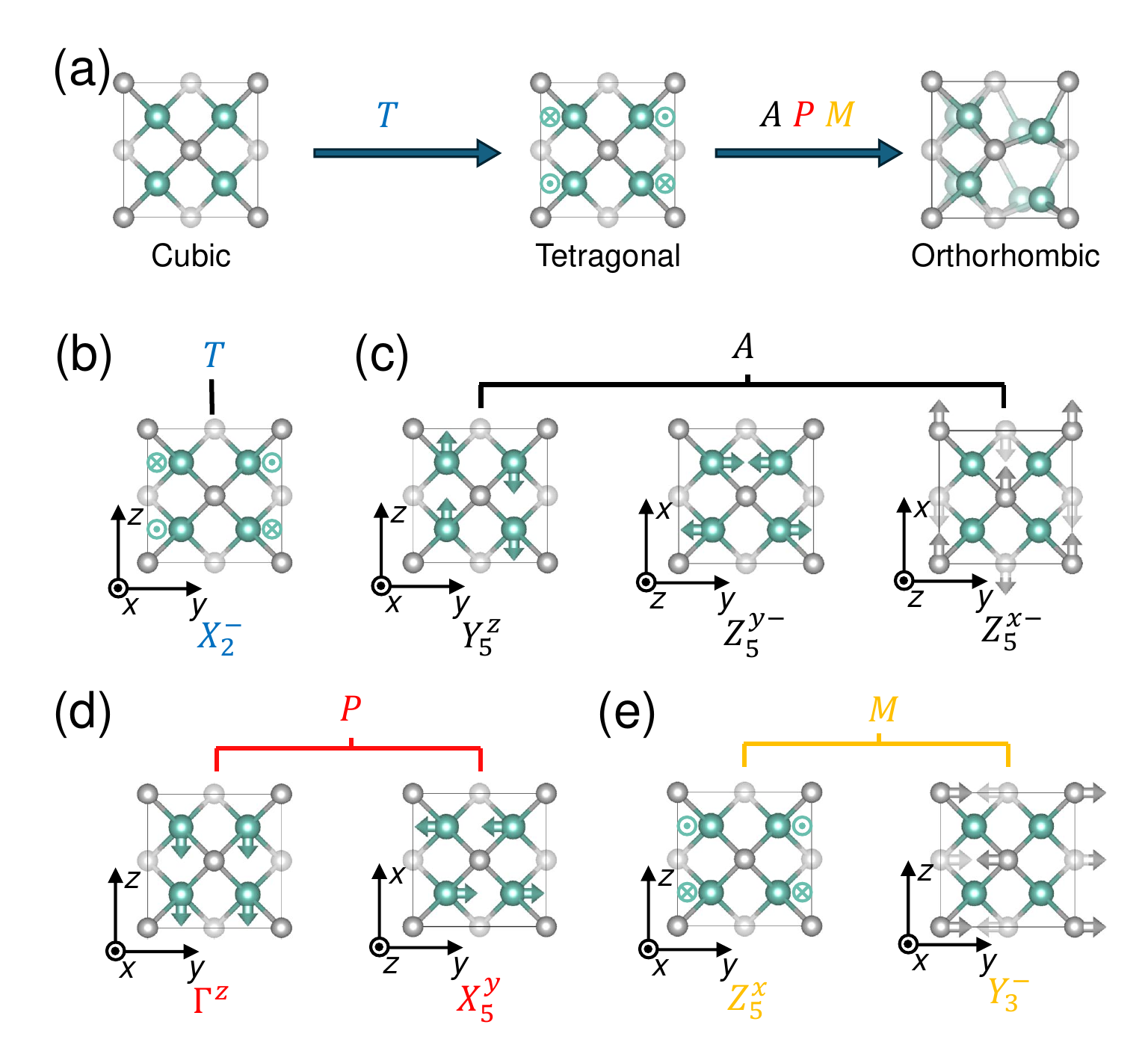}
  \caption{The phases and order parameters of hafnia. (a) the cubic, tetragonal, and orthorhombic phases of hafnia, and the corresponding order parameters involved in the transitions. The grey and green spheres represent Hf and O atoms, respectively. The fainter spheres are in the back of the solid spheres. (b)-(e), the relationship between eight modes defined on cubic symmetry and the order parameters. The first step transition from cubic to tetragonal phase only involves $X_2^-$ modes, which defines order parameter $T$. The other three order parameters $A$ (c), $P$ (d), and $M$ (e) involved in the tetragonal to orthorhombic phase transition are defined on tetragonal symmetry, and thus they are linear combinations of the modes defined in cubic symmetry. Note that $Y_5^z$, $\Gamma^z$, and $Z_5^x$ are the major components of each order parameter.}
  \label{figureS1}
\end{center}
\end{figure}

\begin{figure}
\begin{center}
  \includegraphics[scale=0.35]{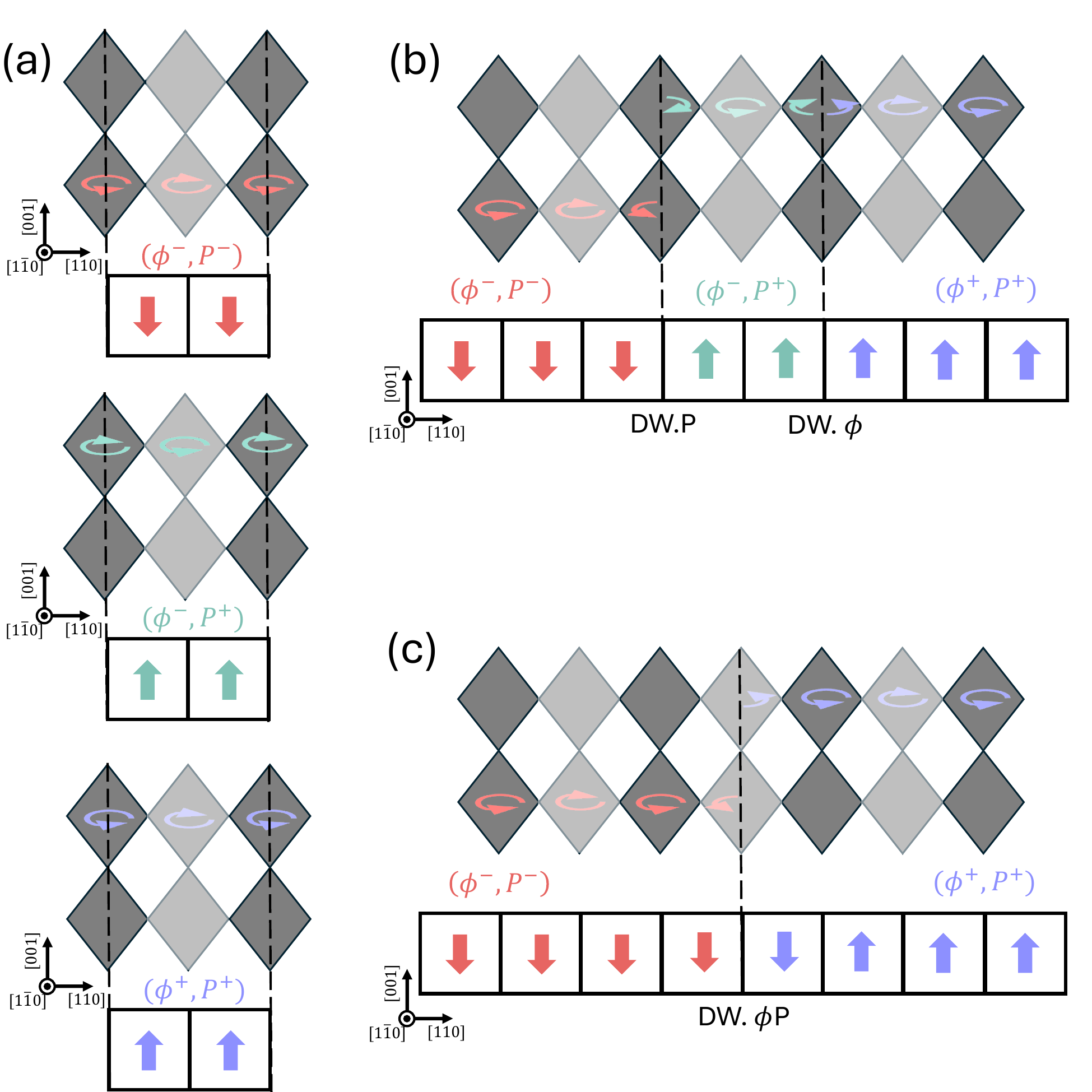}
  \caption{The structure and evolution of the domain wall synthesis reaction in improper ferroelectric perovskite PbTiO$_3$/SrTiO$_3$ superlattice. (a) the structure of domain ($\phi^-,P^-$), ($\phi^-,P^+$), and ($\phi^+,P^+$), respectively. The representation of blocks and colored polarization in the bottom row corresponds to the Fig.5(c) of the main text. Each unit cell of the superlattice contains four oxygen octahedral cages (black diamonds), forming two vertical polarization layers. Thus, each unit cell contains two black blocks. The fainter cages are further behind than the solid cages. The signs of the order parameters (which define the domains of different color) determine which cages are rotated and in what direction. (b) The structure of DW.P and DW.$\phi$ separated by intermediate domain of one unit cell width. The domain wall positions are marked by the disruption of octahedral rotation. The representation in the bottom corresponds to the Fig.5(c)\textcircled{2} of the main text. (c) The structure of DW.$\phi$P in the center of original intermediate domain, generated from synthesis reaction between DW.P and DW.$\phi$. The merging of these two domain walls are completed by changing the rotation pattern in the intermediate domain. The representation in the bottom corresponds to the Fig.5(c)\textcircled{3} of the main text.}
  \label{figureS2}
\end{center}
\end{figure}

\clearpage
\bibliography{apssamp}